\documentclass{ws-ijmpa}
\usepackage{graphicx}
\usepackage{epstopdf}
\usepackage{subfigure}
\usepackage{textcomp}

\def\bkR{{\rm I\kern-.17em R}}
\def\bkC{{\rm \kern.24em \vrule width.05em height1.4ex depth-.05ex \kern-.26em C}}

\def\NAT{{\it Nature} }

\def\PR{{\it Phys. Rev.} }

\def\be{\beta}

\def\bra#1{\langle{#1}|}
\def\ket#1{|{#1}\rangle}

\def\frac#1#2{{\textstyle{{#1}\over {#2}}}}
\def\laq{\raise 0.4 ex \hbox{$<$}\kern -0.8 em\lower 0.62 ex\hbox{$\sim$}}
\def\gaq{\raise 0.4 ex \hbox{$>$}\kern -0.7 em\lower 0.62 ex\hbox{$\sim$}}

\def\be{\begin{equation}}
\def\ee{\end{equation}}
\def\ba{\begin{eqnarray}}
\def\ea{\end{eqnarray}}

\def\dalemb#1#2{{\vbox{\hrule height.#2pt
        \hbox{\vrule width.#2pt height#1pt \kern#1pt \vrule width.#2pt}
        \hrule height.#2pt}}}

\def\dalemb#1#2{{\vbox{\hrule height.#2pt
        \hbox{\vrule width.#2pt height#1pt \kern#1pt \vrule width.#2pt}
        \hrule height.#2pt}}}

\def\gtorder{\mathrel{\raise.3ex\hbox{$>$}\mkern-14mu
             \lower0.6ex\hbox{$\sim$}}}
\def\ltorder{\mathrel{\raise.3ex\hbox{$<$}\mkern-14mu
             \lower0.6ex\hbox{$\sim$}}}

\begin{document}


\title{\bf NONCOMMUTATIVE QUANTUM MECHANICS AND QUANTUM COSMOLOGY\footnote{Based on a talk given by OB in the Second Workshop on Quantum Gravity and Noncommutative Geometry, 22nd-24th September, Universidade Lus\'ofona de Humanidades e Tecnologias, Lisboa}}

\author{CATARINA BASTOS\footnote{Departamento de .F\'\i sica, Instituto Superior T\'ecnico, Avenida Rovisco Pais 1, 1049-001 Lisboa, Portugal. Also at Instituto de Plasmas e Fus\~ao Nuclear, IST. cbastos@fisica.ist.utl.pt}}

\author{ORFEU BERTOLAMI\footnote{Departamento de F\'\i sica, Instituto Superior T\'ecnico, Avenida Rovisco Pais 1, 1049-001 Lisboa, Portugal. Also at Instituto de Plasmas e Fus\~ao Nuclear, IST. orfeu@cosmos.ist.utl.pt}}

\author{NUNO COSTA DIAS\footnote{Departamento de Matem\'{a}tica, Universidade Lus\'ofona de
Humanidades e Tecnologias, Avenida Campo Grande, 376, 1749-024 Lisboa, Portugal. Also at Grupo de F\'{\i}sica Matem\'atica, UL, Avenida Prof. Gama Pinto 2, 1649-003, Lisboa, Portugal. ncdias@meo.pt}}

\author{JO\~AO NUNO PRATA\footnote{Departamento de Matem\'{a}tica, Universidade Lus\'ofona de Humanidades e Tecnologias, Avenida Campo Grande, 376, 1749-024 Lisboa, Portugal. Also at Grupo de F\'{\i}sica Matem\'atica, UL, Avenida Prof. Gama Pinto 2, 1649-003, Lisboa, Portugal. joao.prata@mail.telepac.pt}}

\maketitle

\begin{history}
\received{Day Month Year}
\revised{Day Month Year}
\end{history}

\begin{abstract}

{We present a phase-space noncommutative version of quantum mechanics and apply this extension to Quantum Cosmology. We motivate this type of noncommutative algebra through the gravitational quantum well (GQW) where the noncommutativity between momenta is shown to be relevant. We also discuss some qualitative features of the GQW such as the Berry phase. In the context of quantum cosmology we consider a Kantowski-Sachs cosmological model and obtain the Wheeler-DeWitt (WDW) equation for the noncommutative system through the ADM formalism and a suitable Seiberg-Witten (SW) map. The WDW equation is explicitly dependent on the noncommutative parameters, $\theta$ and $\eta$. We obtain numerical solutions of the noncommutative WDW equation for different values of the noncommutative parameters. We conclude that the noncommutativity in the momenta sector leads to a damped wave function implying that this type of noncommmutativity can be relevant for a selection of possible initial states for the universe.}

\end{abstract}

\section{Introduction}

It is believed that noncommutative spacetime is a fundamental ingredient of quantum gravity\cite{Connes}. Noncommutativity also arises in superstring/M-theory, when describing the excitations of open strings in the presence of a Neveu-Schwarz constant background field\cite{Seiberg}. These ideas have sparked the interest in noncommutative field theories (QFT) [see e.g. [3] and Refs. therein]. In the one-particle sector of QFT, several noncommutative versions of quantum mechanics (NCQM) have been discussed\cite{Gamboa,Zhang,Bertolami1,Acatrinei,Bertolami2,Bastos}.

Most of these models are based on canonical extensions of the Heisenberg-Weyl algebra \cite{Bertolami1,Acatrinei,Bertolami2,Bastos}. In these, time is required to be a commutative parameter and the theory lives in a $2d$-dimensional phase-space of operators with noncommuting position and momentum variables. The extended Heisenberg-Weyl algebra reads:
\be\label{eq1.1}
\left[\hat q_i, \hat q_j \right] = i\theta_{ij} \hspace{0.2 cm}, \hspace{0.2 cm} \left[\hat q_i, \hat p_j \right] = i \hbar \delta_{ij} \hspace{0.2 cm},
\hspace{0.2 cm} \left[\hat p_i, \hat p_j \right] = i \eta_{ij} \hspace{0.2 cm},  \hspace{0.2 cm} i,j= 1, ... ,d
\ee
where $\eta_{ij}$ and $\theta_{ij}$ are antisymmetric real constant ($d \times d$) matrices and $\delta_{ij}$ is the identity matrix. This canonical extension of Heisenberg-Weyl algebra is related to the standard Heisenberg algebra:
\be\label{eq1.3}
\left[\hat R_i, \hat R_j \right] = 0 \hspace{0.2 cm}, \hspace{0.2 cm} \left[\hat R_i, \hat \Pi_j \right]
= i \hbar \delta_{ij} \hspace{0.2 cm}, \hspace{0.2 cm} \left[\hat \Pi_i, \hat \Pi_j \right] = 0 \hspace{0.2 cm},
\hspace{0.2 cm} i,j= 1, ... ,d ~,
\ee
through a class of linear (non-canonical) transformations:
\be\label{eq1.4}
\hat q_i = \hat q_i \left(\hat R_j , \hat \Pi_j \right) \hspace{0.2 cm},\hspace{0.2 cm}
\hat p_i = \hat p_i \left(\hat R_j , \hat \Pi_j \right)
\ee
which are often referred to as Seiberg-Witten (SW) map\cite{Seiberg}. With these transformations, one is able to convert a noncommutative system into a modified commutative system, which is dependent on the noncommutative parameters and of the particular SW map. The states of the system are then wave functions of the ordinary Hilbert space and the dynamics is determined by the usual Schr\"{o}dinger equation with a modified $\eta,\theta$-dependent Hamiltonian. Moreover, one can show that the physical predictions (such as expectation values, probabilities and eigenvalues of operators) do not depend on the particular choice of the SW map\cite{Bastos}.

In the first part of this paper we study some properties of phase-space noncommutative quantum mechanics. We focus on the simple example of the gravitational quantum well (CQW). This is a system of a particle of mass m, moving in the $x'y'$ plane, subjected to the constant Earth's gravitational field, $\mathbf{g}=-g\mathbf{e_{x'}}$, with a horizontal mirror placed at $x'=0$. The measurement of the first quantum states of the GQW for ultra-cold neutrons\cite{Nesvizhevsky} has motivated the study of the noncommutative extensions of the theory\cite{Bertolami1,Bertolami2}.
The aim is to compare the theoretical predictions for the noncommutative system with the available experimental data and hopefully to determine bounds for the values of the noncommutative parameters. Here we will also use this simple system to show explicitly that different choices of the SW map do not affect the physical predictions. To study these issues we will use a qualitative physical feature of the noncommutative gravitational quantum well (NCGQW), the Berry's phase\cite{Bastos1}.

It is well known that quantum states submitted to an adiabatic change can acquire a geometric phase, the Berry's phase. When the system's Hamiltonian is real, this phase can only take values 0 or $\pi$, at the end of a closed path, that is, a non-degenerated wave function must come back to itself or to minus itself. In general, the Berry phase vanishes, but when some point of degeneracy is enclosed in the loop, it can be non-trivial. For instance, Hamiltonians depending on external parameters can be affected by this phase. If the external parameter is classic, typically an external field, the Berry's phase is observed following a non-trivial loop in the space of parameters and carrying through some type of interference between the previous state and the one after completing the closed path. This might be relevant given that the next generation of experiments involving the GQW aim precisely at detecting the transition and interference of states\cite{GranitII}. We will see in the next section that phase space noncommutativity may lead to some momentum shift, which is analogous to an external magnetic field. Thus, one may wonder whether a non-trivial Berry phase may appear in the NCGQW\cite{Bastos1}.

In the second part of this contribution we will study the phase-space noncommutative extension of the quantum Kantowski-Sachs (KS) minisuperspace model\cite{Bastos2}. The KS minisuperspace model has been previously examined in the context of noncommutativity in the configuration space\cite{Compean,Barbosa}. Here we will see that it is the momenta noncommutativity that has the most profound effect on the structure of the theory.

We will determine the noncommutative WDW equation and use the SW map to transform it into a modified commutative equation. The resulting second order, partial differential equation is then further transform into an ordinary differential equation, by taking into account a constant of motion that commutes with the Hamiltonian. Then, this last equation is solved numerically. By examining the resulting physical solutions one is able to restrict the set of possible values of the noncommutative parameters. Moreover, we conclude that it is the momentum space noncommutativity that leads to the richest structure of states. In fact, the fundamental solutions of the WDW equation (which are featureless oscillations for both the commutative and configuration noncommutative cases) display a damping behaviour, suggesting a criterion for selection of states for the early universe.

Before we close this introduction let us mention that various attempts have been considered in order to incorporate gravity into a noncommutative framework. Besides the already mentioned quantum gravity approaches\cite{Connes,Seiberg}, it is relevant to mention attempts to tackle the non-associativity problem that gravity in a noncommutative spacetime background brings in\cite{Bertolami3,Harkumar}, as well as the issues of breaking of Lorentz invariance\cite{Carroll,Bertolami4} and of the stability of fields in a curved spacetime\cite{Bertolami5}.

\section{Noncommutative Gavitational Quantum Well and the Berry Phase}

The gravitational quantum well is described by the following Hamiltonian,
\be\label{eq1.5}
H={{p'_x}^2\over{2m}}+{{p'_y}^2\over{2m}}+mgx'~,
\ee
where $m$ is the neutron mass and $x'$ its height from the mirror, $p'_x$, $p'_y$ the momenta in the $x'$ and $y'$ direction, respectively\cite{Bertolami1}. The solution to this problem, the wave function, can be separated in two distinct parts, corresponding to coordinates $x'$ or $y'$\cite{Bertolami1}. In the $y'$ direction, the wave function corresponds to a group of plane waves with a continuous energy spectrum. In the $x'$ direction, the eigenfunctions can be expressed in terms of the Airy function, $\phi(z)$:
\be \label{eq1.6}
\psi_n(x')=A_n\phi(z)~.
\ee
The roots $\alpha_n$ of the Airy function determine the system's eigenvalues,
\be \label{eq1.7}
E_n=-\bigg({mg^2\hbar^2\over2}\bigg)^{1/3}\alpha_n~.
\ee
The variable $z$ is related to the height $x'$ through the expression,
\be \label{eq1.8}
z=\bigg({2m^2g\over\hbar^2}\bigg)^{1/3}\bigg(x'-{E_n\over mg}\bigg)~.
\ee
The normalization factor for the n-th eigenstate is given by:
\be \label{eq1.9}
A_n=\Bigg[\bigg({\hbar^2\over 2m^2g}\bigg)^{1\over3}\int_{\alpha_n}^{+\infty}dz\phi^2(z)\Bigg]^{-{1\over2}}~.
\ee

Now, let us consider a noncommutative phase space, whose algebra is given by Eq. (\ref{eq1.1}) and $d=2$:
\be\label{eq1.10}
\lbrack x,y]=i\theta\hspace{0.2cm},\hspace{0.2cm}\lbrack p_x,p_y]=i\eta\hspace{0.2cm},\hspace{0.2cm}\lbrack x_i,p_j]=i{\hbar}_{eff}\delta_{ij}\qquad i=1,2~,
\ee
where $\hbar_{eff}=\hbar(1+{\theta\eta}/{4{\hbar}^2})$\cite{Bertolami1}. The Hamiltonian for the noncommutative extension of the GQW, the noncommutative gravitational quantum well (NCGQW), can be obtained using a SW map that allows to re-write Hamiltonian (\ref{eq1.5}) in terms of noncommutative variables. For the following SW map\cite{Bertolami1}
\ba \label{eq1.11}
x'=C\left(x+{\theta\over2\hbar}p_y\right)\hspace{0.5cm},\hspace{0.5cm}y'=C\left(y-{\theta\over2\hbar}p_x\right)~,\nonumber\\
p'_x=C\left(p_x-{\eta\over2\hbar}y\right)\hspace{0.5cm},\hspace{0.5cm}p'_y=C\left(p_y+{\eta\over2\hbar}x\right)~.
\ea
the noncommutative Hamiltonian is given by\cite{Bertolami1}:
\be \label{eq1.12}
H_{NC}^{(1)}={\bar{p_x}^2\over2m}+{\bar{p_y}^2\over2m}+mgx+{C\eta\over2m\hbar}(x\bar{p_y}-y\bar{p_x})+{C^2\over8m\hbar^2}\eta^2(x^2+y^2)~,
\ee
where
\be \label{eq1.13}
\bar{p_x}\equiv Cp_x\hspace{0.2cm},\hspace{0.2cm}\bar{p_y}\equiv Cp_y+{{m^2g\theta}\over{2\hbar}}~,
\ee
with $C\equiv(1-\xi)^{-1}$ and $\xi\equiv \theta\eta/4\hbar^2$.

To first order in the noncommutative parameters $\theta$ and $\eta$ the noncommutative Hamiltonian differs from the
commutative one only by a term proportional to $\eta$\cite{Bertolami1}. So, one can conclude that at leading order
only momenta noncommutativity affects the energy spectrum of the GQW.

Clearly, other SW maps could be used. For instance, one could consider instead the following transformation
\ba \label{eq1.14}
x'=A\left(x+{\theta\over\hbar}p_y\right)\hspace{0.5cm},\hspace{0.5cm}y'=y~,\nonumber\\
p'_y=A\left(p_y-{\eta\over\hbar}x\right)\hspace{0.5cm},\hspace{0.5cm}p'_x=p_x~,
\ea
with $A=(1-4\xi)^{-1}$. The noncommutative Hamiltonian is now given by:
\be \label{eq1.15}
H_{NC}^{(2)}={{p_x}^2\over2m}+{\bar{p}_{y(2)}^2\over2m}+mgx+{A\eta\over m\hbar}\bar{p}_{y(2)}x+{A^2\over2m\hbar^2}\eta^2x^2~,
\ee
where
\be \label{eq1.16}
\bar{p}_{y(2)}=Ap_y+{m^2g\theta\over\hbar}~.
\ee
Thus, one obtains different Hamiltonians for different SW maps. The first three terms in Hamiltonians (\ref{eq1.12}) and (\ref{eq1.15}) correspond to the commutative one, and the remaining terms are interaction like terms that, for instance,
affect the computation of the Berry phase.

>From the identification of the first two energy eigenstates\cite{Nesvizhevsky} it can be shown that the typical momentum scale is bounded by\cite{Bertolami1}:
\be \label{eq1.17}
\sqrt{\eta}\leq0.8\ \mathrm{meV}/c~.
\ee
Assuming that $\sqrt{\theta}\leq1\ \mbox{fm}$, the typical neutron scale, it follows that, $\xi\simeq O(10^{-24})$ and hence that the correction to Planck's constant is irrelevant.

To compute the Berry's phase one uses the following expression\cite{Bastos1}
\be \label{eq1.18}
\gamma_n(\partial S)=i\int_{S}{\sum_{m\neq n}\left[{\bra n\nabla H\ket m\times\bra m\nabla H\ket n}\over{(E_n-E_m)^2} \right]\cdot{d}^2{\bf s}}~.
\ee

Given that the energy spectrum of the GQW is non-degenerate, the Berry phase should vanish. In fact this is verified when we explicitly compute it for the commutative Hamiltonian of the GQW. Now, if we consider the noncommutative Hamiltonian (\ref{eq1.12}) we obtain, for a segment $S$ of the configuration space a phase
\be\label{eq1.19}
\Delta\gamma(S)\sim{\eta}^3~.
\ee
So, the noncommutative parameter associated to momenta noncommutativity, $\eta$, plays a role even in the computation of the Berry's phase for the NCGQW. However, for a full contour $S$, we get a vanishing phase. The same conclusion could be drawn had we considered the alternative noncommutative Hamiltonian (\ref{eq1.15}). Thus different SW maps lead to the same physical phase, and they do not introduce degeneracies in the spectrum of the NCGQW.

\section{The cosmological model}

Let us now turn to quantum cosmology. We consider a cosmological model described by the KS metric. In the Misner parametrization, the line element can be written as\cite{Ryan}
\be\label{eq2.1}
ds^2=-N^2dt^2+e^{2\sqrt{3}\beta}dr^2+e^{-2\sqrt{3}\beta}e^{-2\sqrt{3}\Omega}(d\theta^2+\sin^2{\theta}d\varphi^2)~,
\ee
where $\beta$ and $\Omega$ are the scale factors and $N$ is the lapse function. It is clear that one needs at least two scale factors in order to impose a noncommutative algebra. The Hamiltonian for this metric is obtained via the ADM formalism\cite{Ryan}:
\be\label{eq2.1a}
H=N{\cal H}=Ne^{\sqrt{3}\beta+2\sqrt{3}\Omega}\left[-{P_{\Omega}^2\over24}+{P_{\beta}^2\over24}-2e^{-2\sqrt{3}\Omega}\right]~,
\ee
where $P_{\Omega}$ and $P_{\beta}$ are the canonical momenta conjugate to $\Omega$ and $\beta$, respectively. One can choose $N=24e^{-\sqrt{3}\beta-2\sqrt{3}\Omega}$ as the results are all gauge independent. This is a particular gauge choice, which is only motivated by technical simplicity. At the quantum level, the treatment is manifestly covariant as the lapse function does not enter at all in the formalism. In the next sections, this will be shown explicitly. The classical and the quantum formulations will be considered separately.

\subsection{The Classical Model}

Classically, the equations of motion for the four variables $\Omega$, $\beta$, $P_{\Omega}$ and $P_{\beta}$ can be obtained from the Poisson's bracket algebra. The commutative and the configuration space noncommutativity cases were discussed previously in Refs. [16,17]. Analytical solutions to classical equations of motion were obtained for the two cases.

Here, once again, one considers a canonical noncommutative extension of the Poisson algebra by imposing a noncommutative relation between the two scale factors, $\Omega$ and $\beta$, and between the two canonical momenta, $P_{\Omega}$ and $P_{\beta}$:
\be\label{eq2.10}
\left\{\Omega,P_{\Omega}\right\}=1\hspace{0.1 cm},\hspace{0.1 cm}\left\{\beta,P_{\beta}\right\}=1\hspace{0.1 cm},\hspace{0.1 cm}\left\{\Omega,\beta\right\}=\theta\hspace{0.1 cm},\hspace{0.1 cm}\left\{P_{\Omega},P_{\beta}\right\}=\eta~.
\ee
In the constraint hypersurface
\be\label{eq2.3a1}
{\cal H}\approx 0~,
\ee
the classical equations of motion for the noncommutative system are given by
\ba\label{eq2.11}
&&\dot{\Omega}=-2P_{\Omega}~,\hspace{0.5cm}(a)\nonumber\\
&&\dot{P_{\Omega}}=2\eta P_{\beta}-96\sqrt{3}e^{-2\sqrt{3}\Omega}~,\hspace{0.5cm}(b)\nonumber\\
&&\dot{\beta}=2P_{\beta}-96\sqrt{3}\theta e^{-2\sqrt{3}\Omega}~,\hspace{0.5cm}(c)\nonumber\\
&&\dot{P_{\beta}}=2\eta P_{\Omega}~.\hspace{0.5cm}(d)
\ea
Due to the entanglement among the four variables, an analytical solution is hard to find. However, it is possible to solve numerically this system, and a constant of motion from Eqs. (\ref{eq2.11}a) and (\ref{eq2.11}d) can be obtained\cite{Bastos2}:
\be\label{eq3.1}
\dot{P_{\beta}}=-\eta(-2P_{\Omega})=-\eta\dot{\Omega}\Rightarrow P_{\beta}+\eta\Omega=C~;
\ee
this will play an important role in solving the phase space noncommutative WDW equation.

\subsection{The Quantum Model}

Here and henceforth, one assumes a system of units where $c=\hbar=G=1$, that is Planck units. Assuming the noncommutative parameters, $\theta$ and $\eta$, as an intrinsic feature of quantum gravity, they are expected to be of order one in Planck units.

For the simplest ordering of operators, one obtains the commutative WDW equation for the wave function of the universe by canonical quantization of the classical Hamiltonian constraint, Eq. (\ref{eq2.3a1}),
\be\label{eq2.2}
\exp{(\sqrt{3} \hat{\beta}+2\sqrt{3} \hat{\Omega})}\left[- \hat P^2_{\Omega}+ \hat P^2_{\beta}-48e^{-2\sqrt{3} \hat{\Omega}}\right]\psi(\Omega,\beta)=0~.
\ee
$\hat P_{\Omega}=-i \frac{\partial }{\partial \Omega}$, $\hat P_{\beta}=-i \frac{\partial }{\partial \beta}$ are the fundamental momentum conjugate operators to $\hat{\Omega} = \Omega$ and $\hat{\beta} = \beta$, respectively. Notice that, as it is usual in quantum cosmology, Eq.(\ref{eq2.2}) depends on the prescribed factor ordering. One chooses the simplest factor ordering. This allows for a direct comparison with the results in the literature.

For the commutative case, the solutions of Eq. (\ref{eq2.2}) are \cite{Compean},
\be\label{eq2.3}
\psi^{\pm}_{\nu}(\Omega,\beta)=e^{\pm i\nu\sqrt{3}\beta}K_{i\nu}(4e^{-\sqrt{3}\Omega})~,
\ee
where $K_{i\nu}$ are modified Bessel functions.

One requires that the coordinates and the canonical momenta do not commute. Thus, the extended Heisenberg-Weyl algebra is,
\be\label{eq2.4}
\left[\hat{\Omega}, \hat{\beta} \right]=i\theta\hspace{0.2 cm},\hspace{0.2 cm}\left[\hat P_{\Omega}, \hat P_{\beta}\right]=i\eta\hspace{0.2 cm},\hspace{0.2 cm}\left[\hat{\Omega}, \hat P_{\Omega}\right]=\left[\hat{\beta},\hat P_{\beta}\right]=i~.
\ee
In order to obtain a representation of this algebra (\ref{eq2.4}), one transforms it into the standard Heisenberg algebra through a suitable SW map \cite{Bastos2}:
\ba\label{eq2.8}
\hat{\Omega} =\lambda \hat{\Omega}_{c}-{\theta\over2\lambda} \hat P_{\beta_c} \hspace{0.2cm} , \hspace{0.2cm} \hat{\beta} = \lambda \hat{\beta}_{c} + {\theta\over2\lambda} \hat P_{\Omega_c}~,\nonumber\\
\hat P_{\Omega}= \mu \hat P_{\Omega_c} + {\eta\over2\mu} \hat{\beta}_{c} \hspace{0.2cm} , \hspace{0.2cm} \hat P_{\beta}=\mu \hat P_{\beta_c}- {\eta\over2\mu} \hat{\Omega}_{c}~,
\ea
where the index $c$ denotes commutative variables, i.e. variables for which $\left[\hat{\Omega}_c, \hat{\beta}_c\right]=\left[\hat P_{\Omega_c}, \hat P_{\beta_c}\right]=0$ and $\left[\hat{\Omega}_c, \hat P_{\Omega_c}\right]=\left[\hat{\beta}_c, \hat P_{\beta_c}\right]=i$. This transformation can be inverted only if:
\be\label{eq3.1a}
\xi \equiv \theta \eta <1.
\ee
In this case, the inverse transformation reads:
\ba\label{eq3.2}
\hat{\Omega}_c={1\over\sqrt{1- \xi}}\left( \mu \hat{\Omega} + {\theta\over2\lambda} \hat P_{\beta}\right) \hspace{0.2cm} , \hspace{0.2cm} \hat{\beta}_c={1\over\sqrt{1-\xi}} \left( \mu \hat{\beta} -{\theta\over2\lambda} \hat P_{\Omega}\right)~,\nonumber\\
\hat P_{\Omega_c}={1\over\sqrt{1-\xi}} \left(\lambda \hat P_{\Omega}-{\eta\over2\mu} \hat{\beta} \right) \hspace{0.2cm} , \hspace{0.2cm} \hat P_{\beta_c}={1\over\sqrt{1-\xi}}\left( \lambda \hat P_{\beta}+{\eta\over2\mu} \hat{\Omega} \right)~.
\ea
A relation between the dimensionless constants $\lambda$ and $\mu$ can be found, substituting the noncommutative variables, expressed in terms of the commutative ones, into the commutation relations (\ref{eq2.4}),
\be\label{eq2.8a}
\left(\lambda\mu\right)^2-\lambda\mu+{\xi\over4}=0\Leftrightarrow\lambda\mu={1+\sqrt{1-\xi}\over2}~.
\ee
Hence, through the transformation Eq. (\ref{eq2.8}), one may regard Eq. (\ref{eq2.4}) as an algebra of operators acting on the usual Hilbert space $L^2(\bkR^2)$. In this representation, the WDW Eq. (\ref{eq2.2}) is deformed into a modified second order partial differential equation, which exhibits an explicit dependence on the noncommutative parameters:
\ba\label{eq2.9}
&&\left[-\left(-i \mu {\partial \over \partial {\Omega_c}}+{\eta\over2\mu}\beta_{c}\right)^2+\left(-i \mu {\partial \over \partial {\beta_c}}-{\eta\over2\mu}\Omega_c\right)^2-48\exp{\left[-2\sqrt{3}\left(\lambda\Omega_c+i{\theta\over2\lambda} {\partial \over \partial {\beta_c}} \right)\right]}\right]\times\nonumber\\
&&\times\psi(\Omega_c,\beta_c)=0~.
\ea
This equation is fairly complex and cannot be solved analytically. However, the noncommutative Hamiltonian constraint Eq. (\ref{eq2.9}), for the chosen operator ordering,  commutes with the noncommutative quantum version of the constant of motion Eq. (\ref{eq3.1}):
\be\label{eq3.3}
\hat{C}=\hat{P_{\beta}}+\eta\hat{\Omega}=\sqrt{1-\xi}\left(\mu \hat{P}_{\beta_c}+{\eta\over2\mu}\hat{\Omega}_c\right)
\ee
This allows one to transform the partial differential Eq. (\ref{eq2.9}) into an ordinary differential equation, which can be then solved numerically \cite{Bastos2}. In what follows we present our main results.

\subsection{Solutions}

We consider now Eq. (\ref{eq2.9}) in detail. Let us define a new operator, $\hat{A}=\frac{\hat{C}}{\sqrt{1-\xi}}$, from Eq. (\ref{eq3.3}). It follows that,
\be\label{eq3.4}
\mu \hat{P}_{\beta_c}+{\eta\over2\mu}\hat{\Omega}_c=\hat{A}~.
\ee

As previously mention, the noncommutative WDW Eq. (\ref{eq2.9}) is very complex and no analytical solution is likely to be found. The strategy is then to transform it into an ordinary differential equation and then solve it  numerically. First, one verifies that the constant of motion, Eq. (\ref{eq3.1}), commutes with the Hamiltonian in the constrained space of states:
\be\label{eq3.7}
\left[\hat{P}_{\beta}+\eta\hat{\Omega},\hat{H}\right]=\left[\hat{P}_{\beta}+\eta\hat{\Omega},-\hat{P}^2_{\Omega}+\hat{P}^2_{\beta}-48e^{-2\sqrt{3}\hat{\Omega}}\right]=0~.
\ee
Then, one looks for solutions of Eq. (\ref{eq2.9}) that are simultaneous eigenstates of the Hamiltonian and of the constraint Eq. (\ref{eq3.4}). Let $\psi_a(\Omega_c,\beta_c)$ be an eigenstate of the operator Eq. (\ref{eq3.4}) with a real eigenvalue, $a \in \bkR$, that is:
\be\label{eq3.8}
\left(-i\mu{\partial\over\partial\beta_c}+{\eta\over2\mu}\Omega_c\right)\psi_a(\Omega_c,\beta_c)=a \psi_a(\Omega_c,\beta_c)~.
\ee
This equation admits the solution:
\be\label{eq3.9}
\psi_a(\Omega_c,\beta_c)=\Re(\Omega_c)\exp{\left[{i\over\mu}\left(a-{\eta\over2\mu}\Omega_c\right)\beta_c\right]}~.
\ee
Substituting the wave function Eq. (\ref{eq3.9}) into Eq. (\ref{eq2.9}) one obtains
\ba\label{eq3.10}
&&{\mu}^2\left[{\Re''\over\Re}-i{\eta\over\mu^2}{\Re'\over\Re}\beta_c-{\eta^2\over4\mu^4}\beta_c^2\right]+i\eta\left[{\Re'\over\Re}-i{\eta\over2\mu^2}\beta_c\right]\beta_c-{\eta^2\over4\mu^2}\beta_c^2+\nonumber\\
&&+\left(a-{\eta\over2\mu}\Omega_c\right)^2-{\eta\over\mu}\left(a-{\eta\over2\mu}\Omega_c\right)\Omega_c+{\eta^2\over4\mu^2}\Omega_c^2+\nonumber\\
&&-48\exp{\left[-2\sqrt{3}\left(\lambda\Omega_c-{\theta\over2\lambda\mu}\left(a-{\eta\over2\mu}\Omega_c\right)\right)\right] }=0~,
\ea
where $\Re'\equiv\frac{d \Re}{ d \Omega_c}$. After some algebraic manipulations, one finally gets
\be\label{eq3.11}
\mu^2\Re''+\left(\eta{\Omega_c\over\mu}-a\right)^2\Re-48\exp{\left[-2\sqrt{3}{\Omega_c\over\mu}+{\sqrt{3}\theta\over\lambda\mu}a\right]}\Re=0~.
\ee
Performing the following change of variables,
\be\label{eq3.12}
z={\Omega_c\over\mu}\hspace{0.2 cm}\rightarrow\hspace{0.2 cm}{d\over dz}=\mu{d\over d\Omega_c}~,
\ee
it finally yields for $\phi(z)\equiv\Re(\Omega_c(z))$
\be\label{eq3.13}
\phi''(z)+\left(\eta z-a\right)^2\phi(z)-48\exp{[-2\sqrt{3}z+{\sqrt{3}\theta\over\lambda\mu}a]}\phi(z)=0~.
\ee
This second order ordinary differential equation can be solved numerically. Its solutions depend only on the eigenvalue $a$ and on the noncommutative parameters $\theta$ and $\eta$.

\begin{figure}
\begin{center}
\subfigure[ ~$\theta=\eta=0$ and $a=0.4$]{\includegraphics[scale=0.6]{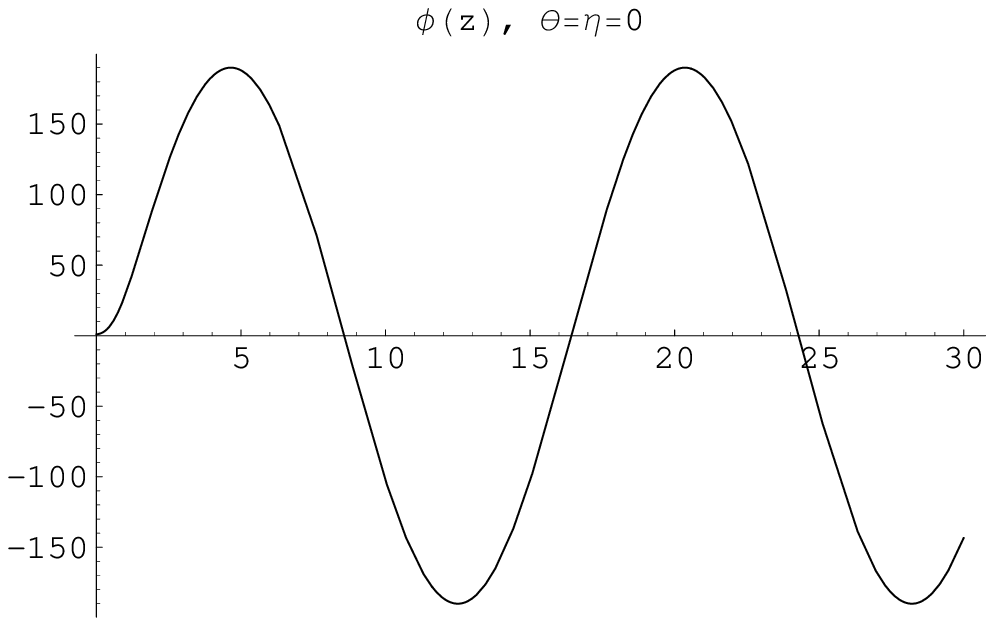}}
\subfigure[ ~$\theta=5$, $\eta=0$ and $a=0.4$]{\includegraphics[scale=0.6]{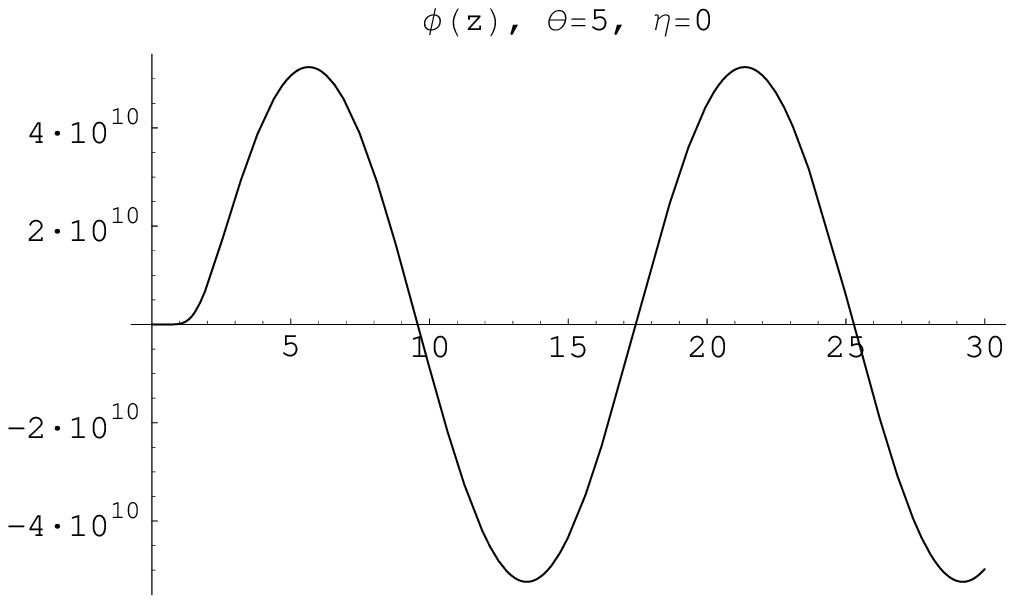}}
\subfigure[ ~$\theta=0$, $\eta=0.1$ and $a=0.565$]{\includegraphics[scale=0.6]{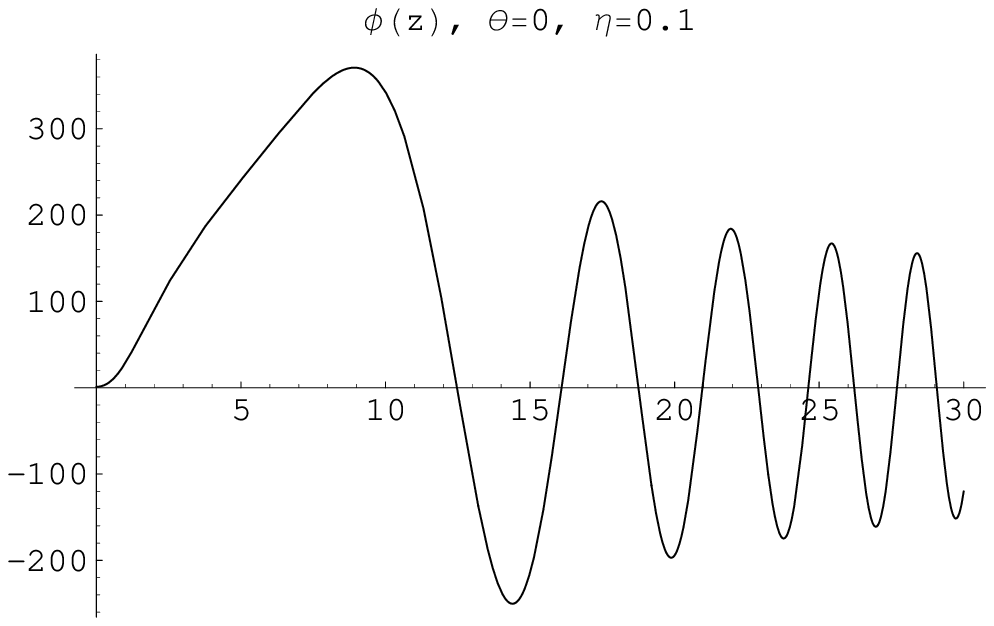}}
\subfigure[ ~$\theta=5$, $\eta=0.1$ and $a=0.799$]{\includegraphics[scale=0.6]{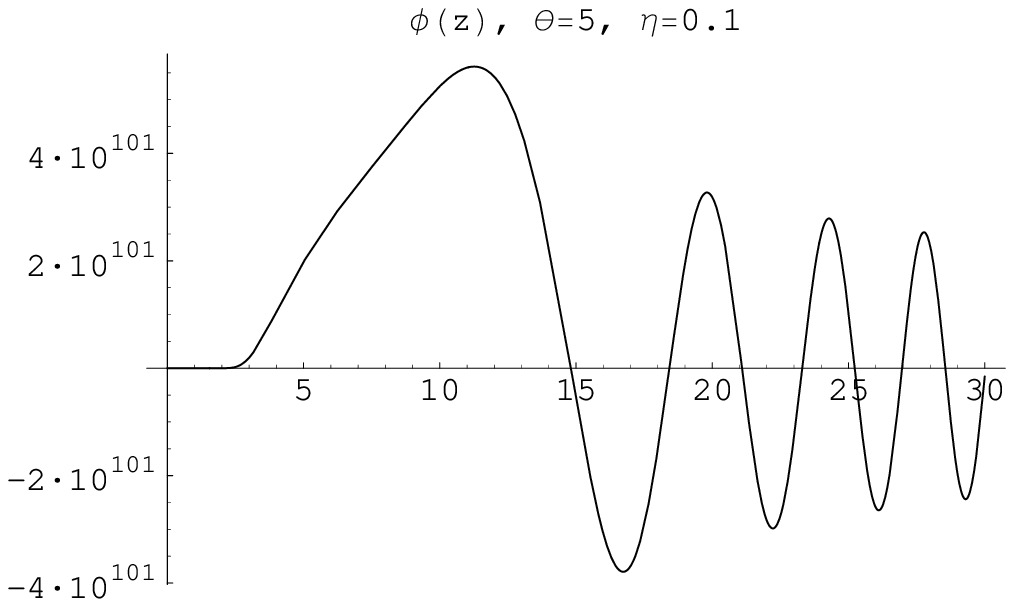}}
\caption{Representation of the numerical solutions of Eq. (\ref{eq3.13}) for different values to the noncommutative parameters. In the four plots $P_{\beta}(0)=0.4$ and $\Omega(0)=1.65$.}
\label{funcaodeonda}
\end{center}
\end{figure}

Our problem involves four initial conditions: $\Omega(0)$, $\beta(0)$, $P_{\Omega}(0)$ and $P_{\beta(0)}$. With the exception of $\beta(0)$, all the others are related to each other due to the Hamiltonian constraint Eq. (\ref{eq2.3a1}). Thus, choosing some numerical values for $P_{\beta(0)}$ and $P_{\Omega}(0)$, one immediately obtains a value for $\Omega(0)$. $\beta(0)$ is an independent initial condition.

The numerical solutions of Eq. (\ref{eq3.13}), for particular sets of values for $a, \theta$ and $\eta$, are depicted in Fig. \ref{funcaodeonda}. The eigenvalue $a$ was taken to be $a=\frac{C}{\sqrt{1-\theta\eta}}$ and is determined through Eq. (\ref{eq3.1}) from the classical values $P_{\beta}(0)$ and $\Omega(0)$ used to generate the solutions of Eqs. (\ref{eq2.11}). These classical values are fairly typical, and once again they were borrowed from the previous studies of the noncommutative KS cosmological model\cite{Barbosa}, so as to allow for comparison with previous results. However, one finds that under variations of the relevant parameters, the qualitative behavior of the obtained wave function is not significantly changed.

Indeed after a thorough analysis of the results, we find that the qualitative features of the solutions displayed in Fig. \ref{funcaodeonda} are unchanged for a rather broad range of values for $\theta,\eta$ and $a$ \cite{Bastos2}. Again, the choice $\theta=5$ is fairly typical in what concerns the properties of the wave function. Furthermore, it is consistent with the point of view that the noncommutative parameters should be of order one close to the fundamental quantum gravity scale. The summary of the main results of Ref. [13] is as follows:

\begin{enumerate}
\item[(i)] For $\theta=5$, the wave function has a damping behavior for $\eta$ in the range $0.05<\eta<0.12$;
\begin{itemize}
	\item For $\eta_c>0.12$  the wave function blows up, suggesting that it is an upper limit for momenta noncommutativity;
\end{itemize}

\item[(ii)] For $\theta>\eta$, varying $\theta$ affects the numerical values of $\phi(z)$, but its qualitative features remain unchanged.
\begin{itemize}
	\item The lower limit for $\eta$ which exhibits a damping behavior is around $\eta\sim0.05$ for all $\theta>\eta$. Clearly, higher $\eta$ values (c.f. Fig. \ref{funcaodeonda}) greatly influence the wave function;
\end{itemize}
\begin{itemize}
	\item For $\eta=0$ the wave function is essentially an oscillation;
\end{itemize}
\begin{itemize}
	\item For $0<\eta<0.05$, the wave function is actually amplified instead of exhibiting a damping behavior;
\end{itemize}

\item[(iii)] For $\eta>\theta$, the damping behavior of the wave function is harder to observe as the wave function does not blow up only for certain values of $\theta$, in particular when $\eta\in[1,2]$;
\begin{itemize}
	\item For $\eta\geq3$ there are no possible range for $\theta$ for which the wave function is well defined;
\end{itemize}

\item[(iv)] For large $z$ the qualitative behavior of the wave function is analogous to the one depicted in Fig. \ref{funcaodeonda} for $z\leq30$.

\end{enumerate}

The criterion to determine bounds for the noncommutative parameters is based on the existence of well defined smooth solutions of the WDW equation. As discussed, solutions do not exist for arbitrary values of $\theta$ and $\eta$. Notice that it is the noncommutativity introduced in the momenta space that affects the behavior of the wave function. This kind of noncommutativity has once again interesting features, as it turns oscillatory functions of the commutative and noncommutative in configuration space cases into damped wave functions. This novel feature is a welcome property as it introduces structure into the wave function, suggesting a natural selection of states for the quantum cosmological model and thus for the set of initial conditions of the classical cosmological model.

\section{Conclusions}

In this contribution we have reviewed the effect of a phase space noncommutativity on the GQW as well as on the minisuperspace KS quantum cosmological model. In both cases, the noncommutativity in the momentum sector is shown to introduce interesting new features. For the NCGQW, the energy spectrum is affected at leading order. The geometric phase through a path in the phase space is dependent on this noncommutative parameter. However, this phase vanishes for a closed path, which indicates that the Hamiltonian of the NCGQW is non-degenerate. This conclusion holds for different SW maps, even though the intermediate calculations yield different results. Thus, different SW maps do not introduce degeneracies and level crossing in the system. This is in agreement with the fact that physical quantities of the system are independent of the chosen SW map\cite{Bastos}.

In the context of quantum cosmology, it has been shown that the noncommutative parameter $\eta$, associated to the momenta, turns
a trivial oscillatory wave function into a damped one. Through the analysis of the noncommutative system, one finds a
classical constant of motion that allows for a numerical solution of the noncommutative WDW equation. The quantum model
is strongly affected by the noncommutativity on the momenta sector, in that the wave function presents a damping behavior
for growing values of the $\Omega$ variable. Thus, the wave function is more peaked for small values of $\Omega$,
which is a rather interesting and a new view of the early universe quantum behavior.

\subsection*{Acknowledgments}

\vspace{0.3cm}

\noindent The work of CB is supported by Funda\c{c}\~{a}o para a Ci\^{e}ncia e a Tecnologia (FCT) under the fellowship SFRH/BD/24058/2005. The work of NCD and JNP was partially supported by Grant No. POCTI/0208/2003 and PTDC/MAT/69635/2006 of the FCT.


\end{document}